\documentclass[11pt,fleqn]{article}

\usepackage{amsfonts}
\usepackage{amsmath}
\usepackage{amssymb}
\usepackage{mathtools}
\usepackage{mathrsfs}
\usepackage{enumerate}
\usepackage{bbm}
\usepackage{graphicx,epsfig,amsthm,color}
\usepackage{pstricks}
\usepackage{subfig}
\usepackage{graphicx}
\usepackage{units}

\usepackage{caption}
\usepackage{booktabs}
\usepackage{textcomp}
\usepackage{array}
\usepackage{cite}
\usepackage{cancel}

\date{\today}

\newcolumntype{z}[1]{>{\RaggedRight\hspace{0pt}}p{#1}}
\newcolumntype{w}[1]{>{\RaggedRight\hspace{0pt}}p{#1}}
\newcolumntype{v}[1]{>{\Centering\hspace{0pt}}p{#1}}

\def \la {\lambda}

\def \a {\alpha}

\def\be{\begin{equation}}
\def\ee{\end{equation}}
\def\bea{\begin{eqnarray}}
\def\eea{\end{eqnarray}}

\def\be{\begin{equation}}
\def\ee{\end{equation}}
\def\bea{\begin{eqnarray}}
\def\eea{\end{eqnarray}}

\def\erp2{{\rm e}^{2\rho}}
\def\erm2{{\rm e}^{-2\rho}}
\def\er4{{\rm e}^{4\rho}}

\def\be{\begin{equation}}
\def\ee{\end{equation}}
\def\bea{\begin{eqnarray}}
\def\eea{\end{eqnarray}}

\def\m0{m_{\nu_{0,i}}}

\def\T0{T_{\nu_0}}

\newcommand{\half}{\frac{1}{2}}

\newcommand{\beqa}{\begin{eqnarray}}
\newcommand{\eeqa}{\end{eqnarray}}
\newcommand{\bpr}{\begin{problem}}
\newcommand{\epr}{\end{problem}}
\newcommand{\bcent}{\begin{center}}
\newcommand{\ecent}{\end{center}}
\newcommand{\bfig}{\begin{figure}}
\newcommand{\efig}{\end{figure}}
\newcommand{\bpc}{\begin{picture}}
\newcommand{\epc}{\end{picture}}

\newcommand{\cam}{{\cal M}}

\newcommand{\nn}{\nonumber}

\renewcommand{\and}{A_{0}^{\nu ,D}(s)}

\newcommand{\bee}{\begin{equation}}

\def\beq{\begin{eqnarray}}
\def\eeq{\end{eqnarray}}

\newcommand{\Ga}{\Gamma}

\newcommand{\bright}{\begin{flushright}}
\newcommand{\eright}{\end{flushright}}
\newcommand{\bminip}{\begin{minipage}}
\newcommand{\eminip}{\end{minipage}}

\DeclareCaptionLabelSeparator{mysep}{\hspace{3pt}:\hspace{3pt}}
\DeclareCaptionLabelFormat{mypiccap}{Fig.\hspace{3pt}{#2}}
\DeclareCaptionLabelFormat{mytabcap}{Tab.\hspace{3pt}{#2}}

\begin{document}

\date{}
\title{
\vskip 2cm {\bf\huge Dilaton stabilization and composite dark matter in the string frame of heterotic-M-theory}\\[0.8cm]}

\author{
{\sc\normalsize Andrea Zanzi\footnote{Email: zanzi@th.physik.uni-bonn.de}\!\!}\\[1cm]
{\normalsize Via de' Pilastri 34, 50121 Firenze - Italy}\\}
 \maketitle \thispagestyle{empty}
\begin{abstract}
{In this paper we further elaborate on our recently proposed solution to the cosmological constant problem - Phys. Rev. D82 (2010) 044006. One of the elements of the solution is the chameleonic behaviour of the Einstein frame dilaton: the mass of the dilaton is an increasing function of the matter density. In that model, a proper structure of the string-frame form factors in the strong coupling region of string theory is assumed. In particular, a stabilizing potential of the string frame dilaton is present and it is supposed to be the result of a quantum calculation. Our main purpose in this article is to point out that the lagrangian of the chameleonic model for the dilaton can be embedded, to a large extent, in heterotic-M-theory. We illustrate some theoretical grounds that support the ansatz about the form factors. In this paper, we break bulk supersymmetry with a massive sterile spinor field (i.e. a bulk neutrino field) and, under certain assumptions about the full M-theory action, we point out the Casimir origin of the stabilizing potential for the string frame dilaton (particular attention is dedicated to the role of the conformal transformations involved). Moreover, we point out the similarity between the string frame dark matter lagrangian and a Ginzburg-Landau lagrangian with sterile neutrino pairs playing the role of dark matter particles. However, more research efforts are necessary to connect the bulk spinor (neutrino) lagrangian with the Ginzburg-Landau lagrangian through direct calculations. A phenomenological test of these ideas is definitely required. \\
}
\end{abstract}

\clearpage

\tableofcontents
\newpage

\setcounter{equation}{0}
\section{Introduction}

In a recent paper \cite{Zanzi:2010rs}, we proposed a stringy solution to the cosmological constant problem \cite{Weinberg:1988cp}. Our model is developed in two different conformal frames, the string frame (S-frame) and the Einstein one (E-frame). As far as the dynamical behaviour of the fields in the S-frame is concerned, in \cite{Zanzi:2010rs} we stabilize all the scalar fields (the dilaton $\phi$ and a scalar field $\Phi$ representative of matter fields). For this reason, the S-frame lagrangian includes a stabilizing potential for the S-frame dilaton \cite{Zanzi:2010rs} which is supposed to be the result of a quantum calculation (basically we assumed a proper structure of the strong coupling form factors). In the S-frame, in the absence of unnatural fine-tunings of the parameters, the cosmological constant is very large. However, once the S-frame scalar fields are stabilized, we perform a conformal transformation to the E-frame (for a detailed discussion of the conformal transformation in this model the reader is referred to \cite{Zanzi:2012du}) and, happily, the E-frame dilaton $\sigma$ parametrizes the amount of scale symmetry of the system keeping under control the E-frame cosmological constant. In other words, in our model, the cosmological constant is under control (and this result is valid including all quantum contributions and without unnatural fine-tunings of the parameters), but only in the E-frame. For this reason we claimed a non-equivalence of different conformal frames at the quantum level \cite{Zanzi:2010rs, Zanzi:2012du}.

Remarkably, in this model, the E-frame dilaton is a chameleon field \cite{Khoury:2003aq, Khoury:2003rn, Mota:2003tc}, namely, it belongs to a group of fields coupled to matter (including the baryonic one) with gravitational (or higher) strength where the interaction matter-chameleon produces an increasing behaviour of the mass of the chameleon as a function of the matter density. On cosmological
distances, where the densities are very small, the chameleons are
ultralight and they can roll on cosmological time scales. On the
Earth, on the contrary, the density is much higher and the field
is massive enough to satisfy all current experimental bounds on deviations from GR. 
In other words, the physical properties of this field
vary with the matter density of the environment and, therefore, it
has been called chameleon. The chameleon mechanism can be considered as a stabilization mechanism. Many other stabilization
mechanisms have been studied for the string dilaton in the
literature. In particular, as far as heterotic string theory is
concerned, we can mention: the racetrack mechanism
\cite{Krasnikov:1987jj, Casas:1990qi}, the inclusion of
non-perturbative corrections to the Kaehler potential
\cite{Casas:1996zi, Binetruy:1996xja, Barreiro:1997rp}, the
inclusion of a downlifting sector \cite{Lowen:2008fm}, Casimir energy \cite{Zanzi:2006xr}.

A chameleonic behaviour of the field, the dilaton, which parametrizes the amount of scale symmetry in the E-frame, led us to a model where scale invariance is abundantly broken {\t locally} (on short distance scales), but it is almost restored {\it globally} (on cosmological distances). There are a number of consequences of this fact, for example: 1) the cosmological constant is under control in the E-frame; 2) the concept of particle is re-examined \cite{Zanzi:2010rs, Zanzi:2012ha} and the string length is chameleonic \cite{Zanzi:2012ha}; 3) matter fields are chameleons \cite{Zanzi:2012ha}. 

At this stage, several points should be further discussed in this chameleonic model for the E-frame dilaton. For example, the theoretical origin of the chameleonic model should be clarified. In particular, some relevant questions are: which one of the various string theories should we choose in order to obtain the lagrangian of the model in a top-down approach? What can we say
about the matching of the chameleonic lagrangian with the stringy lagrangian, namely, what about the non-minimal coupling term, the kinetic term for the dilaton, the stabilizing S-frame potential for the dilaton, the matter part of the lagrangian and the couplings?
In this paper we will partially solve these problems.

We point out that heterotic-M-theory \cite{Lukas:1997fg, Lukas:1998yy, Lukas:1998tt} is able to produce (to a large extent) the required S-frame lagrangian in a top-down approach. Therefore, in the framework of heterotic-M-theory, we briefly touch upon a supersymmetric extension \cite{Brax:2002nt} of the Randall-Sundrum model with two branes \cite{Randall:1999ee} and a bulk dilaton field (for an introduction the reader is referred to \cite{Ovrut:2002hi, Erdmenger:2009ll, Brax:2004xh}). In this model it is possible to render supersymmetric the bulk and the branes separately. When no fields other than the bulk scalar field are present and no supersymmetry (SUSY) breaking effects are present, then the bulk equations agree with junction conditions and the two branes are in a no-force BPS condition \cite{Palma:2004et, Palma:2004fh} where 1/2 of the local SUSY is preserved on the branes. However, our intention in this paper is to leave the BPS configuration. As already pointed out in \cite{Lukas:1997fg}, heterotic-M-theory is constructed so far as an expansion in powers of the 11D gravitational coupling $k^2$ and the full M-theory action is not known. In this paper we are going to take advantage of this fact and we are going to assume that the full M-theory action, namely the action including all higher order corrections in $k^2$, satisfies the following requirements: (1) the non-minimal coupling term and the kinetic term for the dilaton remain formally untouched by these higher order corrections; (2) the corrections produce a de Sitter (dS) detuning for the branes with a resulting shift of the classical background solution from the BPS configuration to a non-BPS non-scale-invariant one; 3) bulk SUSY breaking can be parametrized with a massive bulk spinor field (a sterile neutrino field, like in \cite{Zanzi:2006xr}). 

As far as the matching of the lagrangian of reference \cite{Zanzi:2010rs} with the heterotic-M-theory one is concerned, our results can be summarized in this way:
\begin{itemize}
\item We create a link between the dilaton of reference \cite{Zanzi:2010rs} and the stringy dilaton of the heterotic-M-theory literature. In particular, in the S-frame, we discuss the matching for the kinetic term of the dilaton and for the non-minimal coupling term.
\item We justify the stabilizing potential for the S-frame dilaton of reference \cite{Zanzi:2010rs} exploiting the Casimir calculation of the bulk spinor field performed in reference \cite{Zanzi:2006xr} (for an introduction to Casimir energy see for example \cite{Kirsten:2001wz}). This issue must be further elaborated. In \cite{Zanzi:2006xr}, we mapped the two branes set-up into a 5D ball through a conformal transformation. In this ''bag-frame'' we performed the Casimir calculation \cite{Zanzi:2006xr} and we assumed that a stabilized modulus in one frame corresponds to a stabilized modulus in a different frame \cite{Zanzi:2006xr}. This assumption might generate problems: on the one hand, the potential presence of a correction term (the so-called ''cocycle'' term) might, in principle, generate different dynamical beahviours of one modulus in different conformal frames and, on the other hand, even if we forget about the cocycle, the stabilization of a modulus in one frame might not guarantee, in general, the stabilization of the modulus in a different conformal frame. In this paper, we discuss these issues and we take advantage of the Casimir calculation of reference \cite{Zanzi:2006xr}: we point out in this article that the S-frame dilaton is stabilized exploiting the bag-frame Casimir calculation of reference \cite{Zanzi:2006xr}. In this analysis, particular attention is dedicated to the role of the conformal transformations involved. 
\item We point out the similarity between (A) the string frame dark matter lagrangian of reference \cite{Zanzi:2010rs} and (B) a Ginzburg-Landau \cite{Ginzburg:1950sr} lagrangian where sterile neutrino pairs of the braneworld model mentioned above are playing the role of dark matter particles. However, more research efforts are necessary to connect the bulk spinor (neutrino) lagrangian with the Ginzburg-Landau lagrangian through direct calculations.
\end{itemize}

About the organization of this paper, in section 2 we recall the S-frame lagrangian of our chameleonic model \cite{Zanzi:2010rs}; in section 3 the braneworld model mentioned above is discussed following \cite{Brax:2002nt, Brax:2000xk}. Section 4 is dedicated to the matching of the two lagrangians, on the one hand, the S-frame lagrangian of \cite{Zanzi:2010rs}, on the other hand, the braneworld {\it non}-BPS lagrangian. In section 5 we touch upon some concluding remarks.

\setcounter{equation}{0}
\section{The chameleonic model}
\label{CC}

In this section we will summarize the S-frame lagrangian of our chameleonic model presented recently in \cite{Zanzi:2010rs, Zanzi:2012du}.

\subsection{The S-frame action}
\label{modello}

Our starting point is the string-frame, low-energy, gravi-dilaton
effective action, to lowest order in the $\a'$ expansion, but
including dilaton-dependent loop (and non-perturbative)
corrections, encoded in a few  ``form factors" $\psi(\phi)$,
$Z(\phi)$, $\alpha{(\phi)}$, $\dots$, and in an effective dilaton
potential $V(\phi)$ (obtained from non-perturbative effects). In
formulas (see for example \cite{Gasperini:2001pc} and references therein):
\bea S &=& -{M_s^{2}\over 2} \int d^{4}x \sqrt{-  g}~
\left[e^{-\psi(\phi)} R+ Z(\phi) \left(\nabla \phi\right)^2 +
{2\over M_s^{2}} V(\phi)\right]
\nonumber \\
&-& {1\over 16 \pi} \int d^{4}x {\sqrt{-  g}~  \over
\alpha{(\phi)}} F_{{\mu\nu}}^{2} + \Ga_{m} (\phi,  g, \rm{matter})
\label{3} \eea Here $M_s^{-1} = \la_s$ is the fundamental
string-length parameter and $F_{\mu\nu}$ is the gauge field
strength of some fundamental grand unified theory (GUT) group ($\a(\phi)$ is the
corresponding gauge coupling). We imagine having already
compactified the extra dimensions and having frozen the corresponding
moduli at the string scale (see however sections \ref{hetmth} and \ref{matching} for a more detailed discussion of the geometrical set-up).

Since the form factors are {\it unknown} in the strong coupling
regime, we are free to {\it assume} that the structure of these
functions in the strong coupling region implies an S-frame
Lagrangian composed of two different parts: 1) a scale-invariant
Lagrangian ${\cal L}_{SI}$. This part of our lagrangian has
already been discussed in the literature by Fujii in references
\cite{Fujii:2002sb, Fujii:2003pa}; 2) a Lagrangian which
explicitly violates scale-invariance ${\cal L}_{SB}$.

In formulas we write:

\beq {\cal L}={\cal L}_{SI} + {\cal L}_{SB}, \label{Ltotale}\eeq where the
scale-invariant Lagrangian is given by:

\begin{equation}
{\cal L}_{\rm SI}=\sqrt{-g}\left( \half \xi\phi^2 R -
    \half\epsilon g^{\mu\nu}\partial_{\mu}\phi\partial_{\nu}\phi -\half g^{\mu\nu}\partial_\mu\Phi \partial_\nu\Phi
    - \frac{1}{4} f \phi^2\Phi^2 - \frac{\lambda_{\Phi}}{4!} \Phi^4
    \right).
\label{bsl1-96}
\end{equation}
$\Phi$ is a scalar field representative of matter fields,
$\epsilon=-1$, $\left( 6+\epsilon\xi^{-1} \right)\equiv
\zeta^{-2}\simeq 1$, $f<0$ and $\lambda_{\Phi}>0$.
One may write also terms like $\phi^3 \Phi$, $\phi \Phi^3$ and
$\phi^4$ which are multiplied by dimensionless couplings. However
we will not include these terms in the lagrangian. In general, the coupling $\phi^2 \Phi^2$ can generate a $\phi^4$ term at loop level, so, one interesting question is whether a $\phi^4$-term in the S-frame clashes with a meV cosmological constant in the E-frame. It seems worthwhile to further illustrate this point. When we perform the conformal transformation, the S-frame $\phi^4$-term is mapped to a $M_p^4$-term in the E-frame and the careful reader may be worried about a potential clash in the E-frame between this $M_p^4$-term and the cosmological constant. However, happily, this is not the case: the renormalized Planck mass is exponentially decreasing in the E-frame as a function of the E-frame dilaton $\sigma$ (see \cite{Zanzi:2012du}). 

To proceed further, let us discuss the symmetry breaking Lagrangian
${\cal L_{SB}}$, which is supposed to contain scale-non-invariant terms,
in particular, a stabilizing (stringy) potential for $\phi$ in the
S-frame. For this reason we write: \beq {\cal L}_{\rm
SB}=-\sqrt{-g} (a \phi^2 + b + c \frac{1}{\phi^2}). \label{SB}
\eeq

Happily, it is possible to satisfy the field equations with
constant values of the fields $\phi$ and $\Phi$ through a proper
choice (but not fine-tuned) values of the parameters
$a, b, c$, maintaining $f<0$ and $\lambda_{\Phi}>0$. We made sure
that $g_s>1$ can be recovered in the equilibrium configuration and
that, consequently, the solution is consistent with the
non-perturbative action that we considered as a starting point.

Here is a possible choice of parameters (in string units) \cite{Zanzi:2012du}:
$f=-2/45$, $\lambda_{\Phi}=0.3$, $a=1$, $b=-\frac{13}{72}$, $c=1/108$, $\zeta=5$. In the equilibrium configuration we have
$\phi_0=\frac{1}{2}$ and $\Phi_0=\frac{1}{3}$.

The 4-dimensional curvature in the S-frame is {\it constant} \cite{Zanzi:2012du}. With our choice of parameters we find a positive curvature, $R\simeq10.1$ (in dimensionless units) \cite{Zanzi:2012du}. Therefore we choose the dS metric as our S-frame metric. 

The reader may wonder whether the parameters are very constrained by the equations or not. The answer is that it is very easy to choose the parameters in order to have constant fields. Let us consider, for example, a different choice: $\phi_0=1/2$, $\Phi_0=1/3$, $f=-4$, $\lambda=27$, $a=4$, $b=-3/2$, $c=1/6$, $\zeta=5$. With these non-fine-tuned parameters we find constant fields, positive curvature in 4D and a positive value of the bracket in formula \ref{SB}. 

\setcounter{equation}{0}
\section{The braneworld model}
\label{hetmth}

As far as the model of the previous section is concerned, one relevant issue is related to the stringy origin of the model.
The first weak point of our approach is related to the absence of a top-down derivation of our form factors. It is worthwhile to point out that in our model, only one sector of the lagrangian is scale invariant in the S-frame. This renders the model very ''flexible'': the form factors do not look particularly constrained. Nevertheless, firmer theoretical grounds would be welcome for the entire S-frame lagrangian. 

In order to better illustrate the stringy nature of our E-frame chameleonic dilaton $\sigma$, we will explicitly create a direct connection between the E-frame chameleonic dilaton $\sigma$ and the existing heterotic-M-theory literature. Remarkably, as already discussed in \cite{Horava:1996ma, Horava:1995qa}, strongly coupled $E_8 \times E_8$ heterotic theory is described at low-energy by an 11-dimensional supergravity (SUGRA) theory with the 11-th dimension compactified on a $S^1/Z_2$ orbifold. It has already been shown that 6 dimensions can be consistently compactified on a Calabi-Yau manifold \cite{Witten:1996qb} with size much smaller than the orbifold. The global picture is a 5-dimensional set up with two boundary branes located at the orbifold fixed point. This scenario provides strong theoretical grounds to the idea of {\it braneworld}: our Universe might be a brane embedded in a higher dimensional space. Let us call this brane the {\it visible} one. The remaining brane, the one placed at the opposite orbifold fixed point, will be called {\it hidden} brane and it provides a natural stringy hidden sector.
In this 5-dimensional heterotic-M-theory framework, two scalar fields are present in the low energy action (i.e. a bi-scalar-tensor theory of gravity). These two fields are parametrizing the extradimensional position of the two branes and one of them is connected to the presence of a bulk (dilaton) field (for a review on this model see \cite{Brax:2004xh}).

\subsection{The BPS configuration}
\label{braneworld}

The bulk action is motivated by supergravity and follows
\cite{Brax:2002nt, Brax:2000xk} and it consists of two terms which describe
gravity and the bulk scalar field ($C$) dynamics:
\begin{equation}
S_{\rm SUGRA} = \frac{1}{2\kappa_5^2} \int d^5 x \sqrt{-g_5}
\left( {\cal R} - \frac{3}{4}\left( (\partial C)^2 + U
\right)\right);
\end{equation}

Further, our setup contains two branes. One of these branes has a
positive tension, the other brane has a negative tension.  They
are described by the action
\begin{eqnarray}
S_{\rm brane 1} &=& -\frac{3}{2\kappa_5^2}\int d^5x \sqrt{-g_5}
U_B
\delta(z_1), \label{b1} \\
S_{\rm brane 2} &=& +\frac{3}{2\kappa_5^2}\int d^5x \sqrt{-g_5}
U_B \delta(z_2) \label{b2}.
\end{eqnarray}
In these expressions, $z_1$ and $z_2$ are the (arbitrary)
positions of the two branes, $U_B$ is the superpotential; $U$, the
bulk potential energy of the scalar field, is given by (BPS
relation)
\begin{equation}\label{bpsrel}
U = \left(\frac{\partial U_B}{\partial C}\right)^2 - U_B^2.
\end{equation}
This specific configuration, when no fields other than the bulk
scalar field are present and when the bulk and brane potentials
are unperturbed (no susy breaking effects), is the BPS
configuration \cite{Palma:2004et}, \cite{Palma:2004fh}. When $U_B$
is the constant potential, the Randall-Sundrum model is recovered
with a bulk cosmological constant $ \Lambda_5=(3/8) U=-(3/8)
U_B^2$.

We will also include the Gibbons--Hawking boundary term for each
brane, namely
\begin{equation}
S_{\rm GH} = \frac{1}{\kappa_5^2}\int d^4 x \sqrt{-g_4} K,
\end{equation}
where $K$ is the extrinsic curvature of the individual branes.

The solution of the system above can be derived from BPS--like
equations of the form
\begin{equation}\label{bpseq}
\frac{a'}{a}=-\frac{U_B}{4},\ C'=\frac{\partial U_B}{\partial C},
\end{equation}
where $'=d/dz$ for a metric of the form
\begin{equation}\label{background}
ds^2 = dz^2 + a^2(z)\eta_{\mu\nu}dx^\mu dx^\nu.
\end{equation}
We will consider an exponential superpotential:
\begin{equation}\label{potential}
U_B=4k e^{\alpha C}.
\end{equation}
The solution for the scale factor is
\begin{equation}\label{scale}
a(z)=(1-4k\alpha^2z)^{\frac{1}{4\alpha^2}},
\end{equation}
while the scalar field solution is
\begin{equation}\label{psi}
C = -\frac{1}{\alpha}\ln\left(1-4k\alpha^2z\right).
\end{equation}
In the $\alpha\to 0$ we retrieve the AdS profile
\begin{equation}
a(z)=e^{-kz}.
\end{equation}
In that case the scalar field decouples and the singular point
(for which the scale factor vanishes) at $z_* = 1/(4k\alpha^2)$ is
removed (singularities in braneworld scenarios are analyzed in
\cite{Brax:2001cx}).

\subsection{Leaving the BPS configuration }
In general, there are many ways to leave the BPS configuration: 1) putting matter on the branes 2) detuning the brane tensions 3) breaking bulk SUSY. In this section we will include these three elements simultaneously in our model.

As already pointed out in \cite{Lukas:1997fg}, heterotic-M-theory is constructed so far as an expansion in powers of the 11D gravitational coupling $k^2$ and the full M-theory action is not known. We are going to take advantage of this fact and we are going to assume that the full M-theory action, namely the action including all higher order corrections in $k^2$, satisfies the following requirements: 1) the non-minimal coupling term and the kinetic term for the dilaton remain formally untouched by these higher order corrections and (2) the corrections produce a dS detuning for the branes with a resulting shift of the classical background solution from the BPS configuration to a non-BPS non-scale-invariant one; 3) bulk SUSY breaking can be parametrized with a massive bulk spinor field (a sterile neutrino field, see \cite{Zanzi:2006xr}).

\subsubsection{Moduli Space Approximation: the gravity sector}

In this paragraph we will discuss the moduli space approximation following \cite{Brax:2002nt}.
Two of the moduli of the system are the brane positions. As long as the BPS configuration is respected the position of the branes is described by two constants.  That is,
in the BPS solution the brane positions are arbitrary.  In the
moduli space approximation, these constants are promoted to scalar fields in connection to matter distribution on
branes (remarkably once matter is included in the branes we are already beyond the BPS configuration).  We denote the position of ("visible") brane\footnote{Do not confuse this field with the $\phi$-field of section 2. The proper matching will be discussed in the following paragraphs.} 1 with $z_1 =
\phi(x^\nu)$ and the position of ("hidden") brane 2 with $z_2 =
\lambda(x^\mu)$. We consider the case where the evolution of the
brane is slow. This means that in constructing the effective
four--dimensional theory we neglect terms like $(\partial
\phi)^3$.

In addition to the brane positions, we need to include the
graviton zero mode, which can be done by replacing $\eta_{\mu\nu}$
with a space--time dependent tensor $g_{\mu\nu}(x^\mu)$. Note that
the moduli space approximation is only a good approximation if the
time--variation of the moduli fields is small.

Let us first consider the SUGRA action. Replacing $\eta_{\mu\nu}$
with $g_{\mu\nu}(x^{\mu})$ in (\ref{background}) we have for the
Ricci scalar ${\cal R} = {\cal R}^{(4)}/a^2 + \tilde {\cal R}$,
where $\tilde {\cal R}$ is the Ricci--scalar of the background
(\ref{background}). We explicitly use the background solution
(\ref{scale}) and (\ref{psi}), so that $\tilde R$ will not
contribute to the low--energy effective action. Also, in this
coordinate system, where the branes move, there is no contribution
from the part of the bulk scalar field. Collecting everything we
therefore have \cite{Brax:2002nt}
\begin{equation}
S_{\rm bulk} = \frac{1}{2\kappa_5^2} \int dz d^4 x a^4
\sqrt{-g_4}\frac{1}{a^2}{\cal R}^{(4)} = \int d^4 x \sqrt{-g_4}
f(\phi,\lambda) {\cal R}^{(4)},
\end{equation}
with
\begin{equation}
f(\phi,\lambda) = \frac{1}{2 \kappa_5^2} \int^{\lambda}_{\phi} dz
a^2 (z).
\end{equation}
Let us analyze the Gibbons--Hawking boundary terms.
First, we construct the normal vectors to the brane:
\begin{equation}
n^\mu = \frac{1}{\sqrt{1+ (\partial \phi)^2/a^2}}\left(
\partial^\mu \phi/a^2,1\right).
\end{equation}

Then the induced metric on each brane is given by
\begin{eqnarray}
g_{\mu\nu}^{\rm ind,1} &=& a^2(\phi) g_{\mu\nu}^4 -
\partial_{\mu}\phi
\partial_{\nu} \phi, \\
g_{\mu\nu}^{\rm ind,2} &=& a^2(\lambda) g_{\mu\nu}^4 -
\partial_{\mu}\lambda
\partial_{\nu} \lambda.
\end{eqnarray}
Thus,
\begin{eqnarray}
\sqrt{-g^{\rm ind,1}} &=& a^4(\phi) \sqrt{-g_4}\left[ 1 -
\frac{1}{2a^2(\phi)}(\partial \phi)^2\right], \\
\sqrt{-g^{\rm ind,2}} &=& a^4(\lambda) \sqrt{-g_4}\left[ 1 -
\frac{1}{2a^2(\lambda)}(\partial \lambda)^2\right].
\end{eqnarray}
So the Gibbons--Hawking boundary terms take the form
\begin{eqnarray}
\frac{1}{\kappa_5^2} \int d^4 x a^4 \sqrt{-g_4}\left[ 1 -
\frac{1}{2a^2(\phi)} (\partial \phi)^2 \right] K, \\
\frac{1}{\kappa_5^2} \int d^4 x a^4 \sqrt{-g_4}\left[ 1 -
\frac{1}{2a^2(\lambda)} (\partial \lambda)^2 \right] K.
\end{eqnarray}
The trace of the extrinsic curvature tensor can be calculated from
\begin{equation}
K = \frac{1}{\sqrt{-g_5}}\partial_\mu \left[ \sqrt{-g_5}n^\mu
\right].
\end{equation}
Neglecting higher order terms this gives
\begin{eqnarray}
K = 4\frac{a'}{a}\left[1 - \frac{(\partial \phi)^2}{4a^2}\right].
\end{eqnarray}
The terms for the second brane can be obtained analogously. Using
the BPS conditions and keeping only the kinetic terms, we get for
the Gibbons--Hawking boundary terms
\begin{eqnarray}
&+&\frac{3}{4\kappa_5^2}\int d^4x \sqrt{-g_4} a^2(\phi) U_B(\phi)
(\partial \phi)^2, \\
&-&\frac{3}{4\kappa_5^2}\int d^4x \sqrt{-g_4} a^2(\lambda)
U_B(\lambda) (\partial \sigma)^2.
\end{eqnarray}
Collecting all terms we find
\begin{equation} S_{\rm MSA} = \int
d^4 x \sqrt{-g_4}\left[ f(\phi,\lambda) {\cal R}^{(4)} +
\frac{3}{4}a^2(\phi)\frac{U_B(\phi)}{\kappa_5^2}(\partial \phi)^2
 - \frac{3}{4}
a^2(\lambda)\frac{U_B}{\kappa_5^2}(\lambda)(\partial \lambda)^2
\right] \,. \end{equation}  We see that, as already mentioned above, for a BPS
system the moduli potential is totally absent. However, we included this paragraph in this section dedicated to the non-BPS configuration, because in the moduli space approximation, the branes' position is parametrized by fields and this is connected with the idea that matter particles are localized on the branes.  To proceed further, we redefine the fields as:
\begin{eqnarray}
\tilde \phi^2 &=& \left(1 - 4k\alpha^2 \phi\right)^{2\beta}, \label{posia1}\\
\tilde \lambda^2 &=& \left(1-4k\alpha^2 \lambda\right)^{2\beta}
\label{posia2},
\end{eqnarray}
with
\begin{equation}
\beta = \frac{2\alpha^2 + 1}{4\alpha^2};
\end{equation}
then, the gravitational sector can be written as
\begin{equation}
S_{\rm MSA} =\frac{1}{2k\kappa_5^2(2\alpha^2 + 1)}\int d^4 x
\sqrt{-g_4}\left[ \left(\tilde\phi^2 - \tilde\lambda^2 \right)
{\cal R}^{(4)} + \frac{6}{2\alpha^2 + 1}\left( (\partial
\tilde\phi)^2 -(\partial \tilde\lambda)^2\right)\right] \,.
\end{equation}

Mixed terms like $(\partial_\mu
\tilde\phi)(\partial^\mu \tilde\lambda)$ can be avoided through two new fields (do not confuse the Ricci scalar ${\cal R}$ with
the new field $R$, the radion field):
\begin{eqnarray}
\tilde \phi &=& Q \cosh R, \label{posib1} \\
\tilde \lambda &=& Q \sinh R \label{posib2}.
\end{eqnarray}

The action in the Einstein frame is written as:
\begin{equation}
S_{\rm EF} = \frac{1}{2k\kappa^2_5(2\alpha^2 + 1)} \int d^4x
\sqrt{-g}\left[ {\cal R} -  \frac{12\alpha^2}{1+2\alpha^2}
\frac{(\partial Q)^2}{Q^2} - \frac{6}{2\alpha^2 + 1}(\partial
R)^2\right] \,.
\end{equation}

\subsubsection{The bulk spinor action}
In this pararaph we are going to insert by-hand a massive sterile Dirac spinor in the bulk, parametrizing bulk SUSY breaking. The SPINOR action takes into account the 5D bulk
fermion with mass $m$ of order the fundamental gravity scale and it is
written in the form \cite{Zanzi:2006xr}:

\begin{equation}
\label{action}
   S = \int\!\mbox{d}^4x\!\int\!\mbox{d}z\,\sqrt{-g_5}
   \left\{ E_a^A \left[ \frac{i}{2}\,\bar\Psi\gamma^a
   (\partial_A-\overleftarrow{\partial_A})\Psi
   + \frac{\omega_{bcA}}{8}\,\bar\Psi \{\gamma^a,\sigma^{bc}\} \Psi
   \right] - m\,\mbox{sgn}(\alpha)\,\bar\Psi\Psi \right\} \,.
\end{equation}
We use capital indices $A,B,\dots$ for objects defined in curved
space, and lower-case indices $a,b,\dots$ for objects defined in
the tangent frame. The matrices $\gamma^a=(\gamma^\mu,i\gamma_5)$
provide a four-dimensional representation of the Dirac matrices in
five-dimensional flat space. The quantity $E_a^A$ is the inverse
vielbein and $\omega_{bcA}$ is the spin connection. The sign change of the mass term under $\alpha\to-\alpha$ is
necessary in order to conserve $\alpha$-parity, as required by the
$Z_2$ orbifold symmetry we impose.

\subsubsection{Moduli Space Approximation: the matter sector}

Let us suppose we put some matter on the branes and, in particular, let us localize Standard Model's fields on
the ''visible'' brane (i.e. the UV brane). When the cosmological evolution of the branes is
analyzed in the absence of supersymmetry breaking potentials, it
has been shown in \cite{Palma:2004et, Palma:2004fh} that the
branes are driven by their matter content: one brane is driven
towards the minimum of $U_B$, while the other brane towards the
maximum. In this way it is possible to achieve a full
stabilization of the system, granted that we choose the
superpotential of the theory in a proper way. As already pointed out, if we don't localize
matter on branes and don't break supersymmetry, then the BPS
configuration is respected, the no-force condition between the
branes is present and the system is static. However, as already pointed out above, we assume that loop corrections produce a detuning of the branes' tensions and that dS geometry is achieved. 

We introduce matter as well
as supersymmetry breaking potentials $V(Q,R)$ and $W(Q,R)$ on each
branes. We begin with the potentials: to first order in the moduli
space approximation we get
\begin{equation}
\int d^4 x \sqrt{-g_4} \left[ a^4(\phi)V(\phi) \right]
\end{equation}
with $a^4(\phi) = \tilde \phi^{4/(1+2\alpha^2)}$. The expression
for a  potential $W$ on the second brane is similar with $a(\phi)$
is replaced by $a(\lambda)$. In the Einstein frame we have \cite{Brax:2002nt}:
\begin{equation}
\int d^4 x\sqrt{-g} Q^{-8\alpha^2/(1+2\alpha^2)}(\cosh
R)^{4/(1+2\alpha^2)} V(Q,R) \equiv \int d^4 x\sqrt{-g} V_{\rm
eff}(Q,R),
\end{equation}
where we have defined
\begin{equation}
V_{\rm eff}(Q,R) = Q^{-8\alpha^2/(1+2\alpha^2)}(\cosh
R)^{4/(1+2\alpha^2)} V(Q,R).
\end{equation}
The expression for $W(Q,R)$ in the Einstein frame is \cite{Brax:2002nt}
\begin{equation}
\int d^4 x\sqrt{-g} Q^{-8\alpha^2/(1+2\alpha^2)}(\sinh
R)^{4/(1+2\alpha^2)} W(Q,R) \equiv \int d^4 x\sqrt{-g} W_{\rm
eff}(Q,R),
\end{equation}
where
\begin{equation}
W_{\rm eff}(Q,R) = Q^{-8\alpha^2/(1+2\alpha^2)}(\sinh
R)^{4/(1+2\alpha^2)} W(Q,R).
\end{equation}
The matter action is
\begin{equation}
S_m^{(1)} = S_m^{(1)}(\Psi_1,g^{ind,1}_{\mu\nu}) \hspace{0.5cm}
{\rm and} \hspace{0.5cm} S_m^{(2)} =
S_m^{(2)}(\Psi_2,g^{ind,2}_{\mu\nu}),
\end{equation}
where $g^{ind}$ denotes the {\it induced} metric on each branes
and $\Psi_i$ the matter fields on each branes. Note that we do not
couple the matter fields $\Psi_i$ to the bulk scalar field, and
thus not to the fields $Q$ and $R$. In going to the Einstein frame
we get
\begin{equation}
S_m^{(1)} = S_m^{(1)}(\Psi_1,A^2(Q,R)g_{\mu\nu}) \hspace{0.5cm}
{\rm and} \hspace{0.5cm} S_m^{(2)} =
S_m^{(2)}(\Psi_2,B^2(Q,R)g_{\mu\nu}),
\end{equation}
In this expression we have used the fact that, in going to the
Einstein frame, the induced metrics on each branes transform with
a different conformal factor, which we denote with $A$ and $B$. The
energy--momentum tensor in the Einstein frame is defined as
\begin{equation}
T_{\mu\nu}^{(1)} = 2 \frac{1}{\sqrt{-g}} \frac{\delta
S_m^{(1)}(\Psi,A^2(Q,R)g)}{\delta g^{\mu\nu}}
\end{equation}
with an analogous definition for the energy--momentum tensor for
matter on the second brane.

Hence, the E-frame action is \cite{Brax:2002nt}
\begin{eqnarray}
S_{\rm EF} &=& \frac{1}{16\pi G} \int d^4x \sqrt{-g}\left[ {\cal
R} -  \frac{12\alpha^2}{1+2\alpha^2} \frac{(\partial Q)^2}{Q^2} -
\frac{6}{2\alpha^2 + 1}(\partial
R)^2\right] \nonumber \\
&-& \int d^4 x\sqrt{-g} (V_{\rm eff}(Q,R)+W_{\rm eff}(Q,R)) \nonumber \\
&+& S_m^{(1)}(\Psi_1,A^2(Q,R)g_{\mu\nu}) +
S_m^{(2)}(\Psi_2,B^2(Q,R)g_{\mu\nu}) .
\end{eqnarray}

We point out that in this braneworld scenario we naturally obtain
a multimetric theory. For
more details on the moduli space approximation see
\cite{Palma:2004et}, \cite{Palma:2004fh}.

It seems noteworthy that in the approach of \cite{Brax:2002nt, Brax:2000xk}, while the radion is a chameleon \cite{Brax:2004ym}, the Q-modulus is not: it
is not possible to achieve the desired competition between the
classical potential ($V_{class}(Q) \sim Q^{-\frac{3}{\beta}}$) and the
coupling to matter ($A(Q) \sim Q^{-\frac{1}{2 \beta}}$).

\setcounter{equation}{0}
\section{Matching the two lagrangians}
\label{matching}

In this section we will connect the lagrangian of section 2 to the non-BPS braneworld action which satisfies our requirements about the higher order loop corrections.

\subsection{The string dilaton: kinetic and non-minimal coupling term}

Let us now come back to the dilaton $\phi$ of section \ref{CC} (see also \cite{Zanzi:2010rs}). It is time to obtain some useful results based on our previous discussion about the braneworld model.
One interesting question is whether the chameleonic dilaton $\sigma$ is really a stringy dilaton or a ''phenomenological'' one.
Are we sure we are considering a strongly coupled string theory in the S-frame? After all, the string coupling is connected also to the volume of the compactified extra-dimensions. To solve these problems, at least partially, in this paragraph we are going to focus our attention on the non-minimal coupling term and on the kinetic term of the dilaton. Happily, an excellent match can be established with the heterotic-M-theory lagrangian of the braneworld scenario discussed above. 

To establish a connection with the model of reference \cite{Zanzi:2010rs}, we consider the following geometrical configuration: we put the hidden brane close to the bulk singularity, namely we write $\tilde \lambda \simeq 0$, while the visible brane is far away from the singularity and we write $\tilde \phi \simeq 1$ . In this regime the model is described by a {\it single}-scalar-tensor theory, where the dilaton plays the role of a scalar partner of the metric\footnote{Interestingly, a small $R$ field can be obtained in a low-redshift Universe through a cosmological attractor \cite{Palma:2005wm}.}. Therefore, we suggest to identify the dilaton $\phi$ of \ref{bsl1-96} with the field $\tilde \phi$ of formula \ref{posia1}. In particular, two conditions must be fulfilled. The first one is
\bea
\xi \phi^2=\frac{1}{kk_5^2 (2 \alpha^2 +1)} \tilde \phi^2,
\label{collego}
\eea 
while the second one comes from the kinetic term and it is
\bea
\sqrt{\frac{1}{2kk_5^2 (2 \alpha^2 +1)}  \frac{6}{1+ 2 \alpha^2}} \times \tilde \phi=\frac{\phi}{\sqrt{2}}.
\label{collego2}
\eea
Equations \ref{collego} and \ref{collego2} can be satisfied simultaneously and, in particular, we choose $1+2 \alpha^2 \simeq 1$ which guarantees $\xi \simeq 1/6$ in agreement with our choice of parameters of section \ref{modello}.
Now, the dimensions of $\phi$ are given by \cite{Zanzi:2012ha}:
\bea
[\phi^2]=\frac{M_p^2}{M_s},
\label{dimensioni}
\eea
therefore, from equations \ref{collego} and \ref{dimensioni}, we can write 
\bea
\tilde \phi^2=\frac{M_p^2}{M_s^2}.
\label{gravity}
\eea

This connection between the dilaton $\phi$ of section 2 and the existing string literature is encouraging, however, a proper theoretical origin of (i) the matter part of the lagrangian \ref{bsl1-96} and of (ii) the stabilizing potential for the S-frame dilaton \ref{SB} is still missing. These points will be further discussed in the following paragraphs.

One more remark is in order. As already mentioned in \cite{Brax:2004zs}, in the braneworld set-up discussed above, we are considering a {\it gauged} supergravity theory. Since the gauging procedure is connected to the presence of fluxes, a word of caution is necessary regarding the compactification manifold. When we switch on fluxes, we are led to
heterotic theory on {\it non}-Calabi-Yau manifold. A detailed analysis of these issues in the braneworld set-up would be certainly welcome when attempting to obtain our chameleonic model of section \ref{CC} in a top-down approach. For
further details on fluxes in heterotic theory the reader is
referred to 
\cite{Gurrieri:2007jg,Gurrieri:2004dt,Brustein:2004xn,Becker:2003yv,LopesCardoso:2003sp, Lowen:2008xh} and related papers. For fluxes and moduli stabilization in heterotic-M-theory see \cite{Correia:2009rz} and related papers.
For an introduction to flux compactifications see \cite{Grana:2005jc}.

\subsection{Casimir origin of the S-frame dilatonic potential}

In \cite{Zanzi:2006xr} we mapped the two dS branes set-up discussed above into a 5D bag. In this ''bag-frame'' we exploited the Casimir energy of the bulk spinor (neutrino) field to stabilize the bag radius. If our intention is to change conformal frame and to say that the modulus corresponding to the bag radius is stable in conformal frames which are different from the bag one, several problems must be faced. As already mentioned in the introduction, on the one hand, the potential presence of a correction term (the so-called ''cocycle'' term) should be taken into account and, on the other hand, in general, a stabilized modulus in one frame might not correspond to a stabilized modulus in a different frame.  In this paragraph we suggest to link the Casimir calculation of \cite{Zanzi:2006xr} to the stabilizing potential for the dilaton $\phi$ of the lagrangian \ref{bsl1-96}. As far as the organization of this paragraph is concerned, we will start summarizing some concepts about the Casimir energy of the bulk neutrino and we will proceed discussing the link between the Casimir energy of the bulk spinor and the stabilizing potential for the S-frame dilaton. Particular attention will be dedicated to the role of the conformal transformations involved.

\subsubsection{The 5-dimensional bag}
The Casimir contribution of the bulk spinor has been calculated in \cite{Zanzi:2006xr}. The calculation in the S-frame is complicated and it is preferable to perform a conformal transformation to a new frame that we will call Bag-frame. After the conformal transformation, the two-branes set-up is mapped into a 5D ball. In this paragraph we describe this conformal transformation following \cite{Zanzi:2006xr}, leaving a discussion of the Casimir energy to the following subsections.

In \cite{Zanzi:2006xr}, the starting point of the calculation of the neutrino Casimir energy is a euclideanized form of the metric \ref{background}. On the
Euclidean section the de
Sitter branes become concentric four spheres \cite{59},
\begin{equation}
ds^2 =dz^2+a^2(z)d\Omega^2_4,
\label{sframeds}
\end{equation}
where $d\Omega^2_4$ is the 4-sphere metric.
The metric is conformal to a cylinder $I\times S^4$ \cite{59,60}. Thus,
\begin{equation}
ds^2 =a^2(z)(dy^2+d\Omega^2_4)
\quad\qquad a(z)=(1-4k\alpha^2 z)^{\frac{1}{4 \alpha^2}}\,,
\label{metric}
\end{equation}
where the coordinates are dimensionful and we defined $dy=\frac{dz}{a(z)}$.
The dimensional length $L$ is given by
\begin{equation}
L(\phi,\lambda)=\int_{\phi}^{\lambda}\frac{dz}
{a(z)}=-\frac{1}{k} \frac{1}{4 \alpha^2 -1} [(1-4k \alpha^2 \lambda)^{1-\frac{1}{4 \alpha^2}}-(1-4k \alpha^2 \phi)^{1-\frac{1}{4 \alpha^2}}].
\end{equation}

Let's consider a second transformation given by
\begin{equation}\label{trconf}
y=-R lnr,
\end{equation}
with $0<r<1$ and $R$ is a constant introduced for dimensional reasons (do not confuse this parameter with the radion field). In this way (\ref{metric}) becomes
\begin{equation}
ds^2 =\frac{a^2(z)}{r^2}(dr'^2 + r'^2 d\Sigma^2) \equiv \beta^2 [dr'^2 + r'^2 d \Sigma^2],
\label{metric2}
\end{equation}
where $d \Sigma^2 \equiv R^{-2} d \Omega_4^2$ is the metric for the 4-sphere of radius one, $\beta^2 \equiv a^2/r^2$ is the total conformal
factor and $r' \equiv Rr$ is the dimensionful radial coordinate ($0<r'<R$). The metric (\ref{metric}) is thus conformally related
to a 5-dimensional generalized cone endowed with 4-sphere as a base: a 5-ball of radius R.

In order to keep track of the geometrical "reconfiguration" of the system under the total conformal
transformation, let's consider the following formulas:
\begin{itemize}
\item $z<z^* \equiv \frac{1}{4k \alpha^2}$ where $z^*$ corresponds to the position of the bulk singularity,
\item $y \equiv -\frac{1}{k} \frac{1}{4 \alpha^2 -1} (1-4k \alpha^2 z)^{1-\frac{1}{4 \alpha^2}}$,
\item $y=-Rlnr$.
\end{itemize}
The first one corresponds to the assumption that the branes cannot fall into the bulk singularity.
We assume that the hidden brane is close to the bulk singularity ($z_{hidden} \sim z^*$) while the visible one
is located far away in the bulk ($z_{vis}<<z_{hidden}$). This point requires some additional remarks. We point out that, when we put matter on the branes, the hidden brane is driven towards the bulk singularity: the smaller is the distance between the hidden brane and the bulk singularity, the smaller is the energy of the system (the hidden brane tension is an increasing function of the brane-singularity distance). For this reason, we assume the hidden brane to be stabilized close to the bulk singularity in the S-frame and basically we will ''forget'' about the radion in the remaining part of this paper. We will dedicate our attention to the dilaton field and to the position of the visible brane.
The remaining formulas modify the branes coordinate
in the following way:
\begin{eqnarray}
z \rightarrow -\infty  \Longleftrightarrow y=0 \Longleftrightarrow r=1,  Visible Brane \\
z \rightarrow z^* \Longleftrightarrow y\rightarrow +\infty \Longleftrightarrow r=0, Singularity.
\end{eqnarray}

Remarkably the total conformal transformation translated the cosmological two-branes
set up into a 5-ball. The ball radius is connected to the moduli $\lambda$ and $\phi$ by
\begin{equation}
\frac{L(\lambda,\phi)}{R}=-ln \epsilon,
\end{equation}
where $\epsilon$ is a dimensionless parameter that is small for well-separated branes.

We will start recalling the SPINOR action and studying the effect of the conformal transformation on
the spinor field $\Psi$. The action is
\begin{equation}\label{action}
   S = \int\!\mbox{d}^4x\!\int\!\mbox{d}z\,\sqrt{-g_5}
   \left\{ E_a^A \left[ \frac{i}{2}\,\bar\Psi\gamma^a
   (\partial_A-\overleftarrow{\partial_A})\Psi
   + \frac{\omega_{bcA}}{8}\,\bar\Psi \{\gamma^a,\sigma^{bc}\} \Psi
   \right] - m\,\mbox{sgn}(\alpha)\,\bar\Psi\Psi \right\} \,.
\end{equation}

In the transformation $g_{AB}=\beta^2 g'_{AB}$ the action is rewritten as
\begin{equation}\label{apice}
   S = \int\!\mbox{d}^5x\,\sqrt{-g'_5}
   \left\{ E_a^{A'} \left[ \frac{i}{2}\,\bar\Psi'\gamma^a
   (\partial_A-\overleftarrow{\partial_A})\Psi'
   + \frac{\omega_{bcA}}{8}\,\bar\Psi' \{\gamma^a,\sigma^{bc}\} \Psi'
   \right] - m'\,\mbox{sgn}(\alpha)\,\bar\Psi'\Psi' \right\} \,.
\end{equation}
where the primed quantities are referred to the ball metric and they are given by
$\Psi'=\beta^{2} \Psi$, $m'=\beta m$ and $E_a^{A'}=\beta E_a^{A}$.
We will now use a more compact notation and, omitting the prime, we write (\ref{apice}) as
\begin{equation}
   S = i \int\!\mbox{d}^5x\ \Psi^* D \Psi
\end{equation}
where $D=\slash \nabla + im$.

To proceed further we need to specify boundary conditions for the bulk fermion.
Since we are interested in a chiral theory on the branes, we choose the "option-L" of \cite{Grossman:1999ra}, namely: all the left handed modes are
odd under orbifold parity. In this way the correct boundary condition is
\begin{eqnarray}
z=\phi :\quad P_-\Psi=0,\\
z=\lambda :\quad P_-\Psi=0,
\end{eqnarray}
where $P_{\pm}=\frac12(1 \pm \gamma_5)$. To complete the specification of the
boundary conditions we remember the existence condition for the operator $D^*$: if $P_-\Psi=0$,
this requires that the normal derivative $(\partial_y - m)P_+ \Psi=0$
should vanish. In summary, we have defined two subspaces in direct sum generated by
the operators $P_{\pm}$; while on the first space (-) we have Dirichlet condition, on the second one (+)
we impose Robin boundary condition. In other words the bulk fermion satisfies mixed boundary conditions (for
further details see \cite{Flachi:2001ke,Moss:1996ry,Moss:2004un}).

In this way our model is the 5-dimensional "extension" of the MIT bag model \cite{Chodos:1974je, Chodos:1974pn}. In 4D quarks and gluons
are free inside the bag, but they are unable to cross the boundary. This last condition
corresponds precisely to the mixed boundary
conditions discussed above: a chiral theory on the branes corresponds to the condition
that no quark current is lost through the boundary. The zeta function for massive fermionic fields inside
the bag has been considered in \cite{Elizalde:1997hx}. Analogous calculations have been developed for massive scalar field 
(\cite{Bordag:1995gs, Bordag:1996ma, Bordag:1995gm}).
Functional determinants were discussed in \cite{Bordag:1995zc, Bordag:1996fw, Dowker:1995sw, Elizalde:1996nb, Elizalde:1996zw, Moss:2004un, Dowker:1996ej}.
For moduli stabilization with zeta function regularization see also \cite{Garriga:2001ar, Flachi:2003bb}.

The 4D QCD analysis of \cite{Elizalde:1997hx} has been applied to the Heterotic-M-theory branes in \cite{Zanzi:2006xr} where the MIT bag model calculation has been extended from four to five dimensions and, happily, the neutrino Casimir energy turned out to be stabilizing: the neutrino Casimir energy as a function of the bag radius $R$ shows an AdS minimum \cite{Zanzi:2006xr}. Our next step is to take advantage of this theoretical calculation in order to stabilize the 4D S-frame dilaton $\phi$ (i.e. to justify the stabilizing potential \ref{bsl1-96}).

\subsubsection{Casimir energy and the effective action}

In order to clarify the connection between Casimir energy and effective action,
we write the Laplace-like operators (related to the fields whose Casimir contribution we want to evaluate) in the following unified form,
namely
\beq P = -g^{\rho \nu} \nabla_\rho \nabla _\nu -E
,\label{eq1.1g} \eeq
where $g^{\rho \nu}$ is the Riemannian metric
of the manifold $\cam$, $\nabla$ is a connection and $E$ an
endomorphism defined on $\cam$. We are confronted with the
task of calculating expressions for the effective action $\Gamma$ of the type
\beq \Gamma [V] =-\frac
1 2 \ln [\mbox{det} (P / \mu^2)],\label{eq1.2} \eeq
where $\mu$ is a mass parameter.

Clearly, expression (\ref{eq1.2}) is not defined because the
eigenvalues $\lambda_n$ of $P$, \beq P\phi_n = \lambda_n \phi_n,
\label{eq1.3} \eeq grow without bound for $n\to \infty$. Of
course, there are various possible regularization procedures; let
us mention only Pauli-Villars, dimensional regularization and zeta
function regularization. In this last case the basic idea is to generalize the identity,
valid for a $(N\times N)$-matrix $P$,
\beq \ln \mbox{det}\,\, P =
\sum_{n=1}^N \ln \lambda_n = - \frac d {ds}  \sum_{n=1}^N
\lambda_n^{-s} |_{s=0} = -\frac d {ds}
         \zeta_P (s) |_{s=0} , \nn
\eeq with the zeta function \beq \zeta_P (s) = \sum_{n=1}^N
\lambda_n^{-s} ,\nn \eeq to the differential operator $P$
appearing in (\ref{eq1.3}) by \beq \ln \mbox{det}\,\, P =
-\zeta_P' (0) , \label{eq1.4} \eeq with
\beq \zeta_P (s) =
\sum_{n=1}^\infty \lambda_n^{-s} .\label{eq1.5} \eeq

The Casimir energy of the bulk neutrino has been analyzed with this formalism in \cite{Zanzi:2006xr} through the spatial part of the Laplace operator $P_s=H^2$, where
H is the Hamilton operator for the Dirac spinor inside the bag.
To see this, write the Hamilton operator formally as \beq H =
\sum_k E_k \left(N_k +\frac 1 2 \right) , \nn \eeq
with $N_k$ the
number operator, to obtain for the vacuum energy
\beq E_{vac} =
<0|H|0>=\frac 1 2 \sum_k E_k .\label{casimir} \eeq

The regularization we considered in \cite{Zanzi:2006xr} for the spinor field is

\beq E_{vac} &=& -\frac {\mu^{2s}} 2 \sum_k
(E_k^2)^{1/2-s} |_{s=0}
     = -\frac{\mu^{2s}} 2 \zeta _{P_s} (s-1/2) |_{s=0} \nn\\
 &=&- \frac 1 2 FP\,\,\zeta _{P_s} (-1/2) -
\frac 1 2 \left( \frac{1}{s} + \ln \mu^2 \right) Res
\zeta _{P_s} (-1/2) \nn\\
&=& -\frac 1 2 FP\,\,\zeta _{P_s} (-1/2)
+ \left( \frac 1 s + \ln \mu^2 \right) \frac 1 {2\sqrt{4\pi}} a_{D/2}(P_s),
\label{eq1.13a}
\eeq
where $FP\,\zeta$ is the finite part of the zeta function.

For a detailed description of the remaining calculations related to the neutrino Casimir energy see \cite{Zanzi:2006xr}.

\subsubsection{The cocycle}

As already mentioned above, conformal tranformations are useful in Casimir calculations: the shift from the S-frame to the bag-frame renders the computation easier. In general, if our intention is to evaluate the S-frame potential, a correction term called cocycle might be present in the effective action (for some calculations, see for example \cite{Moss:2003zk, Moss:2004un}). The cocycle function is basically a conformal anomaly term and in this paragraph we are going to briefly touch upon its definition.

The computation of the cocycle function is heavily dependent on knowing appropriate
heat kernel coefficients. These heat kernel coefficients are defined in $D$
dimensions by an asymptotic expansion,
\begin{equation}\label{heatker}
{\rm tr}\left(F e^{-t P }\right)\sim \sum_{n=0,1/2,1,...}^\infty
a_n[P,F] t^{n-\frac{D}{2}}.
\end{equation}
In five dimensions we require $a_{5/2}[P,F]$, which can be found in the
literature \cite{Kirsten:2001wz}.

Given a sequence of metrics $g_\epsilon=e^{-2\epsilon F}g$, and operators
$P_\epsilon$, it can be shown that \cite{Kirsten:2001wz}
\begin{equation}\label{cociclo}
W[P(\epsilon),P]:= \zeta'_{\epsilon}(0)-\zeta'_{0}(0)=2 \int_0^{\epsilon} d{\tau}
a_{D/2}[P(\tau),F].
\end{equation}
The relevant conformal transformation is (\ref{metric2})
with boundaries at $z_1=\phi$ and $z_2=\lambda$.

Following \cite{Garriga:2001ar} we can write
\begin{equation}
    \int_0^\epsilon    a_{5/2}(f_{\tau},P_{\tau}) d\tau
    ={d\over d  D} a^{D}_{5/2}(P_\epsilon)\Big|_{D=5}-
       {d\over d  D} a^{D}_{5/2}(P_0)\Big|_{D=5}.
\end{equation}
Thus, the cocycle, namely the integral in (\ref{cociclo}),  can
be evaluated in two different ways: (a)
by using the explicit expression of $a_{5/2}(F,P)$
given by Kirsten \cite{Kirsten:2001wz} and (b) by taking the derivative of the
coefficients $a^D_{5/2}(P_\epsilon)$
with respect to the dimension.

\subsubsection{Casimir energy and the S-frame potential}

First of all, the 5D bulk frame \ref{sframeds} can be identified with the S-frame in heterotic-M-theory. The geometry of the 4-dimensional Universe is de Sitter and our moduli corresponds to a strongly coupled theory in the S-frame (see also formula \ref{gravity} with $\tilde \phi \simeq 1$). The potential we would like to obtain from the Casimir energy is $Z_2$-symmetric and it can be written as 
\bea V=a\phi^2+b+\frac{c}{\phi^2}.\label{fidep}\eea The presence of a $\phi^4$-term does not alter our argument. We start analyzing the action in the S-frame. In our model, we are considering heterotic theory in the strong coupling region. The 5D effective action of heterotic-M-theory has been discussed in \cite{Lukas:1998yy, Lukas:1998tt, Lukas:1997fg}. In their derivation the authors exploited a reduction of the $Ho\check{r}ava-Witten$ theory on a Calabi-Yau space. They showed that the resulting 5-dimensional theory is a gauged SUGRA.
The background solution is BPS and there always exists a background solution with two moduli representing the interbrane distance and the dilaton field. This is the theoretical origin of the model described in section \ref{braneworld}. In this scenario loop corrections to the effective potential have been calculated (see for example the 1-loop calculation of \cite{Moss:2004un, Garriga:2001ar}) considering the BPS solution
as a classical background solution. A word of caution is necessary. As already pointed out in the introduction, heterotic-M-theory has been constructed so far as an expansion in powers of the 11D gravitational coupling $k^2$ and the full M-theory action is not known \cite{Lukas:1997fg}. In this paper we leave the BPS background exploiting higher order corrections. This point requires some additional remarks. We assume that the full M-theory action, namely the action including all higher order corrections in $k^2$, satisfies these requirements: 1) the non-minimal coupling term and the kinetic term for the dilaton remain formally untouched by these higher order corrections; (2) the corrections produce a dS detuning for the branes with a resulting shift of the classical background solution from the BPS configuration to a non-BPS non-scale-invariant one; 3) bulk SUSY breaking can be parametrized with a massive bulk spinor field (a sterile neutrino field, see \cite{Zanzi:2006xr}).  

There are a number of consequences of this method: 

A) First of all, in this way, both branes have dS geometry like in the bag-frame calculation of reference \cite{Zanzi:2006xr}, where the bag radius is stabilized through Casimir energy. We follow the same approach in this paper. In other words, we consider the 5D bag of reference \cite{Zanzi:2006xr} with a stabilized radius. The question we will analyze in this section is whether this stable bag configuration guarantees a fixed position of the visible brane in the S-frame. To answer the question, we will analyze in this section the conformal transformation from the bag frame to the S-frame.

B) When we come back from the bag-frame to the S-frame, one problem may arise in connection to the cocycle term but, happily, the cocycle is absent in the set of conformal transformations we are interested in. This point we mentioned last needs some comments. The cocycle function is basically a conformal anomaly and it contributes to the quantum effective action after a conformal transformation has been performed. In our model however, our action is a non-perturbative heterotic-M-theory action, where the contribution of loops (at all orders) have been taken into account. The classical background solution we are considering is a non-BPS configuration with detuned branes in the dS direction such that scale invariance is absent. Hence, in our model, we assume we are equipped with a non-scale-invariant background solution which we choose as a starting point of our quantum Casimir calculation. We infer that there is no conformal anomaly in this 5D model, because the action which we consider formally as the source of our classical background solution is not scale invariant. Summarizing, when we move from the S-frame to another frame (and back) through a conformal transformation, the cocycle function is absent in our model. 

In the 5D bag frame, considering the stabilizing behaviour of the neutrino Casimir energy \cite{Zanzi:2006xr} (notice that we evaluated only the $j=1/2$ contribution \cite{Zanzi:2006xr}), we form a bound state of neutrinos. To a certain extent, the situation is reminiscent of quark condensation in 4D and also gaugino condensation in supersymmetric theories. Hence, we are led to the following scenario:
\begin{itemize}
\item A bound state of sterile neutrinos is formed in 5D through (strongly coupled) gravitational interaction.
\item The energy scale at which gravitational interaction gets strongly coupled ($\alpha_G \simeq 1$) is planckian (remember that in 4D $\alpha_G \equiv 16 \pi G p^2$, where $p$ is the momentum transfer). 
\item The Planck mass in the bag-frame is generated through neutrino condensation, it is RG-invariant and it fixes the mass scale of the Casimir energy.
\item This bag-frame RG-invariance in 4D can be also interpreted as a z-independence (i.e. a moduli-independence) in 5D through AdS/CFT correspondence \cite{Maldacena:1997re}. 
\end{itemize}

Summarizing, we are considering the stable bag with a constant Planck mass and then we move to the S-frame.
When we come back to the S-frame, another problem must be faced before we can say that the S-frame dilaton is stabilized through Casimir energy (i.e. that formula \ref{fidep} can be considered the relevant part of the lagrangian close to the Casimir minimum): when we stabilize the dilaton in the bag frame we cannot say that the dilaton is stabilized in another frame (even if the cocycle term is absent), because the potential presence of two different dynamical behaviours of the dilaton in two conformally-related frames must be considered. Let us discuss this point more carefully focusing our attention on the bag frame and the S-frame. As already mentioned above, the Planck mass is RG-invariant in the bag-frame and the RG-invariance in 4D can be interpreted as z-independence in 5D. Moreover, this constant planckian scale of the bag-frame is the mass scale which enters into the conformal transformation linking the two frames (the bag-frame and the string one). Consequently, there is a one-to-one map in 4D between the vacua of the bag frame and the vacua of the S-frame: one bag frame vacuum is mapped into one single S-frame vacuum, because the Planck mass of the relevant conformal transformation is constant. Since we are considering a quantum theory, the fields involved in the conformal transformation must receive an expectation value (renormalization must be taken into account, see also \cite{Zanzi:2012du}) and we know that, in the bag-frame, the dilaton is stabilized and the renormalized Planck mass is constant. From these elements we gathered, we infer that the S-frame dilaton is stabilized through the Casimir contribution.  As far as the cosmological constant is concerned, we imagine that the non-fine-tuned detuning in the dS direction due to higher order corrections is related also to the $b$-parameter of formula \ref{SB}.

Interestingly, once the S-frame dilaton is stabilized, we know that a chameleonic behaviour of the E-frame dilaton is compatible with an exponential dependence of the E-frame Planck mass on $\sigma$ just after the conformal transformation to the E-frame \cite{Zanzi:2012du}. Fixing the S-frame 4d dilaton induces an exponential dependence of the 4d E-frame Planck mass on the E-frame chameleon $\sigma$ and the reader may be worried by a non-RG-invariant Planck mass. We point out however, that only the E-frame is physical and, therefore, there is no clash between an RG-invariant bag-frame Planck mass and an RG-non-invariant E-frame Planck mass. The chameleonic E-frame dilaton $\sigma$ is parametrizing the energy scale of the system and the RG-running of the Planck mass can be reformulated through its $\sigma$-dependence.

\subsection{The (dark) matter lagrangian}

The purpose of this paragraph is to find a (partial) motivation for the structure of the S-frame matter lagrangian of section 2, namely
\begin{equation}
{\cal L}_{\rm matter}=\sqrt{-g}\left(  -\half g^{\mu\nu}\partial_\mu\Phi \partial_\nu\Phi
    - \frac{1}{4} f \phi^2\Phi^2 - \frac{\lambda_{\Phi}}{4!} \Phi^4
    \right).
\label{materia}
\end{equation}

Let us start considering once again the bulk massive spinor. We know that it is sterile, namely it interacts only gravitationally. Interestingly, the formation of gravity-induced neutrino pairs is reminiscent of electron-electron pairs in superconductivity. Now we further elaborate on this point we mentioned last following \cite{Ryder:1985wq} (see also \cite{Kittel, Strocchi:1985cf} and chapter 21 of reference \cite{Weinberg:1996kr}). 

It is common knowledge that superconductivity is a phenomenon present in many metals where the absence of resistance is shown at very low temperatures. Hence, in these metals there can be persistent currents which screen out the magnetic flux (Meissner effect). From a particle physics perspective, the phenomenon is the result of a spontaneous breaking of electromagnetic gauge invariance in the material. Consequently, let us discuss an example of spontaneous symmetry breaking starting from the following lagrangian for a scalar field $\theta$ (written in the static case $\partial_0 \theta=0$):
\bea
{\cal L}=-({\bf \bigtriangledown} -ie{\bf A}) \theta \cdot  (\bigtriangledown + ie {\bf A}) \theta^* - m^2 \mid \theta \mid ^2-\lambda \mid \theta\mid ^4 - \frac{1}{2} (\bigtriangledown \times {\bf A})^2,
\label{superc}
\eea
where ${\bf A}$ is a vector potential.
$-{\cal L}$ is the so-called Ginzburg-Landau free-energy \cite{Ginzburg:1950sr} and we have $m^2 \propto (T-T_c)$ near the critical temperature $T_c$.

The use of the $\theta$ field finds its theoretical grounds in the Bardeen-Cooper-Schrieffer (BCS) theory \cite{Bardeen:1957mv} where, under certain conditions, there is an attractive force between electrons which is responsible for the formation of electron-electron pairs. These bosonic pairs fall into the same state at low temperatures (Bose-Einstein condensation) and, hence, the scalar field $\theta$ can be exploited in the description of the macroscopic system.
When $T>T_c$, we have $m^2>0$ and the minimum of the free energy is at $\mid \theta\mid =0$. On the contrary, when $T<T_c$, we have $m^2<0$, the minimum of the free energy is at
\bea
\mid \theta\mid ^2=-\frac{m^2}{2\lambda}>0
\eea
and the symmetry is spontaneously broken. The mass of the $\theta$ field fixes the inverse of the length scale (known as the correlation length) where the variations in $\theta$ take place. 

Let us come back to the bulk spinor. We point out in this paper the similarity between the S-frame matter lagrangian \ref{materia} and the Ginzburg-Landau lagrangian. In particular, consider the following identifications: $\theta=\Phi$, $m^2=f \phi^2/4$ and $\lambda=\lambda_{\Phi}/4!$. We parametrize the fraction of condensed neutrinos through the order parameter $\theta$. Now one question: what kind of matter field is this neutrino-neutrino pair? If we remember that these condensing neutrinos are sterile, we infer that it is not possible to identify the $\theta$ field with standard model's matter. However, we can identify the $\theta$ field with a Dark Matter (gauge neutral) field. In this way, we are led to the following set up. The Casimir energy of the bulk spinor brings the visible brane towards a stable configuration which corresponds to a non-vanishing number density of the extradimensional neutrinos. The bosonic $\nu-\nu$ condensate is induced by gravitational attraction in 5D and, at low energies, its 4D description is given by a composite scalar field which plays the role of bosonic Dark Matter. The mass of the Dark Matter field in the S-frame might be dilaton-dependent, because the position of the visible brane fixes the UV cut-off scale, namely the S-frame Planck mass (see also the Gherghetta's lectures in \cite{Kazakov:2006kp}). It seems worthwhile to point out that in this model we have two scalar degrees of freedom, on the one hand, the distance between the two branes, on the other hand, the position of the center of mass of the branes. For this reason, once the hidden brane is kept close to the bulk singularity, another degree of freedom is available, namely, the position of the visible brane. In the S-frame, the larger is the value of the dilatonic z-coordinate of the visible brane, the smaller is the distance between the two branes, the smaller is the 4D UV planckian cut-off and the lower is the energy scale where gravity becomes strongly coupled. Hence, a smaller $\mid  m^2 \mid $ might be expected when we imagine to move the visible brane towards the bulk singularity. In this way, the S-frame $\phi^2 \Phi^2$ term might be justified (exploiting formula \ref{collego} we see that $\phi$ goes to zero for colliding branes). 

It would be rewarding to develop a direct calculation in order to connect the bulk neutrino lagrangian with the Ginzburg-Landau one.
Remarkably, in the minimum of the potential we have $\phi \simeq \Phi$. 
The lagrangian for standard model's matter will be discussed in a future work.

\section{Concluding remarks}

In this paper we further elaborated on our recently proposed solution to the cosmological constant problem \cite{Zanzi:2010rs, Zanzi:2012du}. In that model the dilaton in the string frame is stabilized through a potential which is supposed to be the result of a quantum calculation. In this article we pointed out that the model of reference \cite{Zanzi:2010rs} can be embedded, to a large extent, in heterotic-M-theory and we partially clarified the origin of its string frame lagrangian.

As already pointed out in the introduction, heterotic-M-theory has been constructed so far as an expansion in powers of the 11D gravitational coupling $k^2$ and the full M-theory action is not known \cite{Lukas:1997fg}. In this paper we assumed that the full M-theory action, namely the action including all higher order corrections in $k^2$, satisfies these requirements: 1) the non-minimal coupling term and the kinetic term for the dilaton remain formally untouched by these higher order corrections; (2) the corrections produce a dS detuning for the branes with a resulting shift of the classical background solution from the BPS configuration to a non-BPS non-scale-invariant one; 3) bulk SUSY breaking can be parametrized with a massive bulk spinor field (a sterile neutrino field, see \cite{Zanzi:2006xr}).  
With these assumptions, the matching of the two lagrangians has been discussed. In particular, we touched upon the matching of the kinetic terms for the dilaton and of the non-minimal coupling terms in the two lagrangians. We suggested to exploit Casimir energy of bulk (neutrino) field to stabilize the string frame dilaton taking advantage of the bag-frame analysis of reference \cite{Zanzi:2006xr}. In this paper we moved from the bag frame to the string one and particular attention has been dedicated to the conformal tranformations involved. Moreover, we suggested to identify dark matter particles with a composite scalar field which represents in 4D the result of extradimensional-neutrino condensation. In this way the dark matter lagrangian in the S-frame might be justified as a Ginzburg-Landau lagrangian.

One word of caution is necessary regarding dilaton stabilization. As already mentioned above only the $j=1/2$ contribution has been evaluated (see \cite{Zanzi:2006xr}). Interestingly, as far as the S-frame vacuum energy is concerned, meV scale is not required in this frame and, hence, the model is able to include quantum corrections easily in the effective action without destroying the solution to the (E-frame) cosmological constant problem.  

Many lines of development are present, for example:\\
A) an interesting line of development will analyze potential connections between our model and the ekpyrotic scenario of \cite{Khoury:2001wf}.\\ 
B) Another project should try to mix a S-frame dilaton running towards the strong coupling regime (see also \cite{Gasperini:2001pc}) with the chameleonic solution to the cosmological constant problem of reference \cite{Zanzi:2010rs}.\\
C) More research efforts are necessary to\\
(C1) study the couplings of the model in detail;\\
(C2) discuss the baryonic sector of the theory and the lagrangian for matter on the hidden brane;\\
(C3) connect the bulk spinor (neutrino) lagrangian with the Ginzburg-Landau lagrangian through direct calculations.\\
D) Needless to say, a careful investigation of the phenomenological aspects of this proposal would be welcome.

\subsection*{Acknowledgements}

Special thanks are due to Gianguido Dall'Agata, Antonio Masiero, Marco Matone, Massimo Pietroni and Roberto Volpato for valuable discussions.


\providecommand{\href}[2]{#2}\begingroup\raggedright\endgroup

\end{document}